# Waving in the rain

Cavaleri Luigi[a]*, Bertotti Luciana[a], Bidlot Jean-Raymond[b]

a) Institute of Marine Sciences, CNR, Arsenale – Tesa 104, Castello 2737/F, 30122 Venice, Italy
b) European Centre for Medium-Range Weather Forecasts, Shinfield Park, Reading, Berkshire, RG2 9AX, U.K.

15 January, 2015

*) :corresponding author:
Luigi Cavaleri
ISMAR
Arsenale – Tesa 104
Castello 2737/F
30122 Venice, Italy
ph +39-041-2407955
fx +39-041-2407940
luigi.cavaleri@ismar.cnr.it

Key points

rain affects wind wave generation and dissipation processes
white-capping tends to disappear under heavy rain
strong relation between wind wave generation and white-capping

**Abstract**

We consider the effect of rain on wind wave generation and dissipation. Rain falling on a wavy surface may have a marked tendency to dampen the shorter waves in the tail of the spectrum, the related range increasing with the rain rate. Following the coupling between meteorological and wave models, we derive that on the whole this should imply stronger wind and higher waves in the most energetic part of the spectrum. This is supported by numerical experiments. However, a verification based on the comparison between operational model results and measured data suggests that the opposite is true. This leads to a keen analysis of the overall process, in particular on the role of the tail of the spectrum in modulating the wind input and the white-capping. We suggest that the relationship between white-capping and generation by wind is deeper and more implicative than presently generally assumed.

# 1 – Wind waves and rain

It is a common experience among seafarers, both of the past and today, that "rain calms the sea". Figure 1 shows such a situation. The irregular surface clearly indicates that we are not dealing with swell. At the same time, for whoever had the chance to witness a stormy sea, the lack of short wave features and breakers is macroscopically evident.

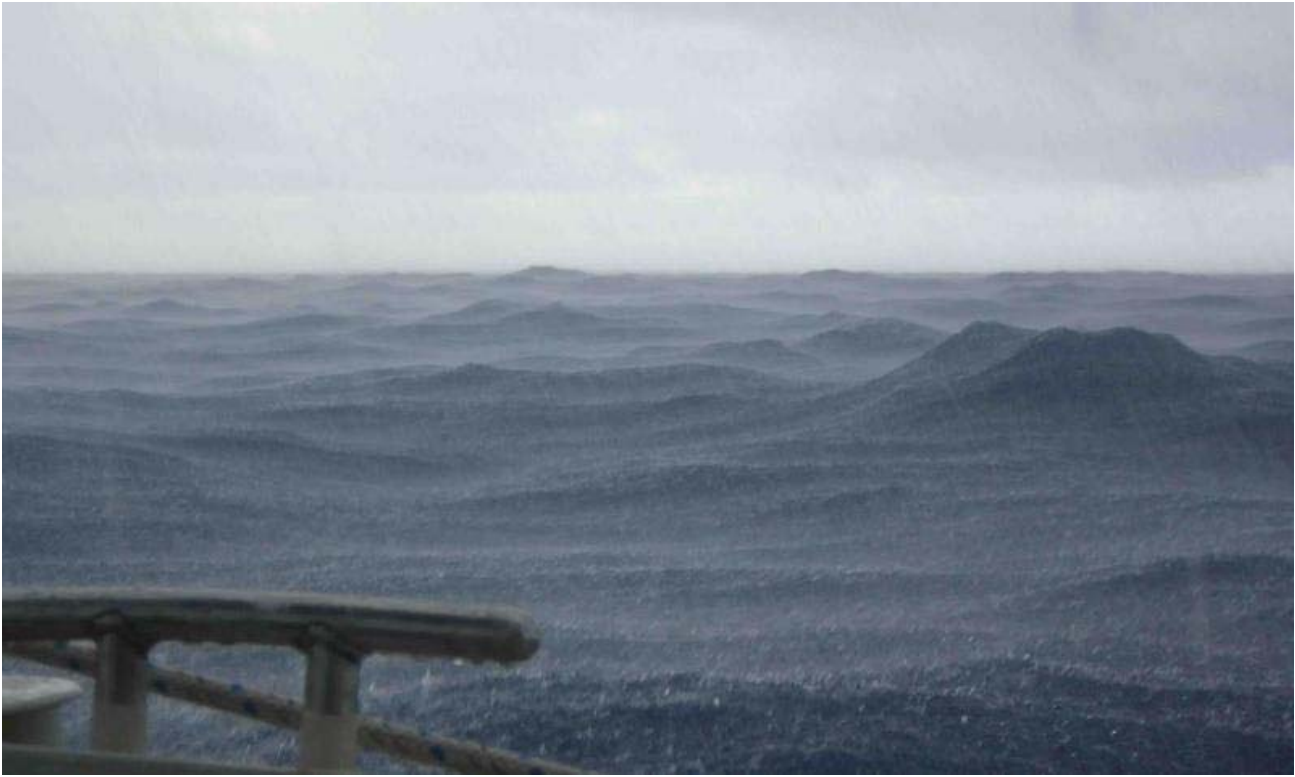

Figure 1 – Wind sea in rainy conditions (courtesy of Ginni Callahan). Environmental conditions (on-board estimate): wind speed 15 $ms^{-1}$, significant wave height > 1 m, mainly wind sea, very heavy rain.

The related processes have attracted scholar attention since long time ago. The first report we managed to trace back is by *Reynolds* [1875] who experimented with artificial drops falling on calm water. He reported that, if sufficiently large and energetic, each drop has a tendency to create a vortex ring and, more in general, to increase turbulence in the upper layer. He concluded saying that this had the potential of attenuating water waves. Reynolds came back to the subject in 1900 with more details, but basically the same idea.

With few exceptions, as *Manton* [1973], the subject laid basically dormant, at least according to our information, for a long while. Wave modeling, entering the digital era in the '70s, was too busy with more fundamental and quantitatively relevant processes to pay attention to rain. The matter came into focus again in the late '80s, early '90s, in connection with the use of remote sensing instruments, scatterometer and Synthetic Aperture Radar (SAR) in particular. Relying for signal detection on the interaction of the emitted radar signal with the centimetric waves at the sea surface, the efficiency of the instruments was obviously depending on rain. This led to a number of studies and reported results. *Tsimplis and Thorpe* [1989], *Tsimplis* [1992], *Beya et al.* [2010] and *Peirson et al.* [2013] studied the effect of artificial rain on mechanically generated waves in a wave flume. *Poon et al.* [1992], as also *Braun et al.* [2002], went a step further using wind generated waves in a flume with rain falling on a limited section of it. With some differences among the various reported results, the emerging general picture is the following. Rain, if intense enough as usually the case in laboratory experiments, leads to a small scale turbulence in the first few centimeters below the water surface. It also increases the surface roughness at the centimetric scale (order of frequency 10 Hz). Witnessing a downpour on a lake or a small pond will provide immediate evidence of the little messy surface. The consequent surface motion is very low, incoherent and, with the exception of an oblique rain component, isotropic. In practical terms it does not contribute directly to wind wave generation.

Longer waves, from a few centimeters upwards, are attenuated. Corresponding evidence for the shorter waves, i.e. the high tail of the spectrum, is given in the various laboratory experiments quoted above. A quantification for the longer waves, not achievable in laboratory conditions, has been provided by *Le Méhauté and Khangaonkar* [1990, henceforth referred to as LMK] who made a keen analysis of the effect of falling rain drops on an underlying wave field, taking into consideration also the possible wind effect, i.e. of rain falling at a marked angle with respect to the vertical. For a 50 $mmh^{-1}$ precipitation rate (soon to be commented about) they derived a 38, 5, 0.5 % hourly wave height decay for 1, 10, 100 meter long waves respectively.

The practical implications for scatterometer and SAR instruments have been well defined, among others, by *Weissman et al.* [2012] and *ESA* [2013]. The Ku band signal (~14 GHz, 2.1 cm) is strongly affected, as it was the case for QuikSCAT. C band (~5.3 GHz, 5.6 cm) seems to fare better, and it has been the preferred choice for ASCAT. *Chen et al.* [1998] used this difference, together with radiometer data, to estimate the rain distribution on the oceans. However, according to ESA the problem is not so well defined because the transition zone between increased and decreased wave heights, certainly in the 5 to 10 cm range, depends on the rain rate, the drop size distribution, the wind speed, and the time history of the rain event.

From the point of view of wave modeling, our present main interest, the seemingly accepted fact is that, when rain is present, for waves from a few centimeters upwards there is a marked attenuation, rain rate dependent and rapidly decreasing with the wave length. The case of oblique rain adding energy and momentum to waves will be discussed in the final Section 6. For our later discussion we will refer to LMK, as all the laboratory experiments deal with short waves and very large rain rates. For a proper perception of the relevance of the process, compare these with the typical 10 $mmh^{-1}$ of the extra-tropical countries (there are local exceptions). The 50 $mmh^{-1}$ rates quoted in LMK and the 40 $mmh^{-1}$ by *Peirson et al.* [2013] (the minimum rate they used in their laboratory experiments) are already closer to local peak values. Nevertheless, the relevant seemingly accepted fact (one of us, L.C., has direct experience in this sense) is that a rainy sea has a smoother appearance than in dry windy conditions. This will be our starting point for the following considerations. These will be developed in the following Section 2, while Sections 3 and 4 will quantify the modeling implications. In Section 5 we verify our work hypothesis with operational data, finding, to our initial surprise, opposite results. The implications and the search for possible explanations are the subject of the last Section 6.

**2 – Rain attenuates waves in the tail of the spectrum**

We start from the assumption in the title of this section, as derived from the above cited literature and experience, and follow the implication for wave modeling under the umbrella of the theories amply applied in wind wave modeling.

Our focus is on wind wave generation. The generally accepted process, at least in operational wave modeling, is the one proposed by Miles in 1957, and later perfected by *Janssen* [1989, 1991]. Granted the decomposition of a wavy surface into a number of spectral sinusoidal components suggested by Pierson and Marks in 1952, Miles' mechanism envisages for each component a smooth similarly wavy air flow that, because of phase shift, ends up inputting energy to the wave. The further step by Janssen was to point out that the energy, and momentum, transferred into waves come, hence should be subtracted, from the blowing air. This must imply a slowdown of the wind field and, as immediate consequence, also lower wave heights. The direct results [*Janssen*, 1989, 1991, 2008] confirmed this approach.

A relevant detail of this process is that (*Janssen*, 2004, page 109) "The main contribution to the wave stress is determined by the medium- to high-frequency gravity waves with dimensionless speed c/u* in the range of 1 to 10, as these are the waves with the highest growth rate". Therefore anything affecting the tail of the spectrum is expected to affect the interaction between atmosphere and ocean, i.e. between wind and waves.

The point of our research is straightforward. Consider a situation in which, for any external reason, the high frequency tail of the wave spectrum is cancelled. This may be because of rain, as we have described in the previous section, or because of oil on the sea surface, as done during the second world war for 'men at sea' recovery in stormy conditions, or because of the mixture of water and ice in close to freezing conditions (grease ice). In this situation the momentum input by wind to waves will be strongly reduced. In turn this will imply stronger wind speeds and, following Miles, higher wave heights. This is our work hypothesis, i.e. that in rainy areas wind and waves should be (marginally) stronger and higher respectively than anticipated by the present common modeling approach.

**3 – Modeling set-up**

To verify our hypothesis, we have used the coupled modeling system operational at the European Centre for Medium-Range Weather Forecasts (ECMWF, Reading, U.K.) till November 2013. The system implies a full two-way coupling between the meteorological and WAM wave models (see respectively, *Simmons and Gibson* [2000], and *Komen et al.* [1994], *Janssen* [2008]).

We have used the T1279 version of the system, i.e. 16 and 28 km spatial resolution respectively for the meteorological and wave global models. The wave model was run with 36 frequency ($f_1$ = 0.0345 Hz) with 1.1 geometric progression, and 36 directions starting at 5 degree clockwise with respect to North. Focused on a period to be soon specified, we have first done a reference run with the standard set-up. The run was then repeated introducing a zero-ing of the wave spectrum, beyond a frequency $f_c$ depending on the rain rate, in the calculation of the high frequency contribution to the surface stress. The rule we adopted was $f_c$*rain=54, with rain (rate) given as mmh$^{-1}$. Whichever the rain rate, the minimum zero-ed frequency was 0.5 Hz. As previously specified, the incoherent centimeter scale surface turbulence due to rain is ignored, also because it becomes appreciable only in heavy rain conditions, when the rain attenuated spectral tail range is larger. Overall this was clearly a rather crude and arbitrary approach. For instance, it is natural to think of a progressive smoothing of the zero-ing with frequency. However, lacking any more precise indication, especially for limited rain rates, our purpose was only to verify the physical principle, leaving, if the results were positive, to a second stage the task to better quantify the details. As the results will show, this was a convenient move.

To verify our hypothesis, obviously we need windy and stormy areas. This excluded the equatorial zone for lack of sustained winds. We also excluded hurricanes and typhoons as isolated and extreme events, not suitable for a first approach. We focused our attention on the North Atlantic Ocean, especially in the 35°-70° latitude range (see Figure 5). Here we have frequent storms and plenty of measured data, both from buoys and satellites.

Searching for a suitable period, we have explored the wind and rain maps from January 2006 till December 2010. Our choice for a full month period was for December 2009. In this month, the ECMWF system was still running with T799 resolution (about 25 km). The archived results are regularly available at 6-hour interval for analysis, 3-hour for forecast. However, rain events can be shorter and spatially localized and measured data are available in the intermediate hours. Therefore we were interested in higher time and space resolution. Hence in this phase we ignored the archived data and we repeated the simulation for the full December 2009 with the already specified T1279 resolution, saving the results at one hour interval. As said above, this was first done with the standard set-up, and then repeated with the zero-ed spectral tail according to rain. Standard meteorological and wave integrated parameters and full 2D spectra were saved at each grid point. Each run was done as a sequence of 24 hour forecasts at 12 hour interval. For each forecast, the 13-24 hour section was retained. This provided a full month simulation suitable for the subsequent analysis.

Measured data were from moored buoy and satellite measurements. A large portion of the buoy data comes from the archive of in-situ data routinely obtained from the global telecommunication system with the rest supplemented as part of the data exchanged under the JCOMM project of model performance inter-comparison [*Bidlot et al.*, 2007]. Both wind and wave data were used. For the satellites we excluded the scatterometer data because of their doubtful use in rainy areas. C band data, e.g. ASCAT, are less affected by rain and regularly used for operational purposes. However, as we will see, we deal with limited differences that would be comparable with the, albeit limited, error of the scatterometer in rainy areas. For similar, although less doubtful, reasons we excluded also the use of altimeter wind data. Altimeter wave height data seem a much more solid information. We used these data from both ENVISAT and Jason-2 satellites. The data were extracted from the RADS database [*Naeije et al.*, 2008]; see http://rads.tudelft.nl/rads/literature.shtml for details. The model results were linearly interpolated in space and time to the buoy and altimeter data.

**4 – Results of the experiments**

We focus our attention on the ten meter wind speed $U_{10}$ and the significant wave height $H_s$ as basic indicators of the performance of the coupled model system.

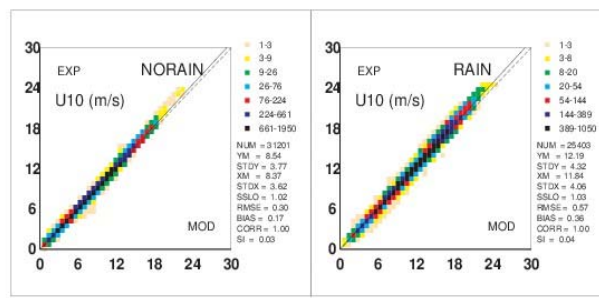

Figure 2 − EXPerimental run with reduced surface roughness under rain versus the regular MODel results. Surface wind speeds over the North Atlantic Ocean (see Figure 5 for the considered area).
Left panel results in the non-rainy areas, right panel in rainy areas. Colors represent, in geometric progression, the number of cases in each pixel (see the side scale). Also the overall statistics is reported.

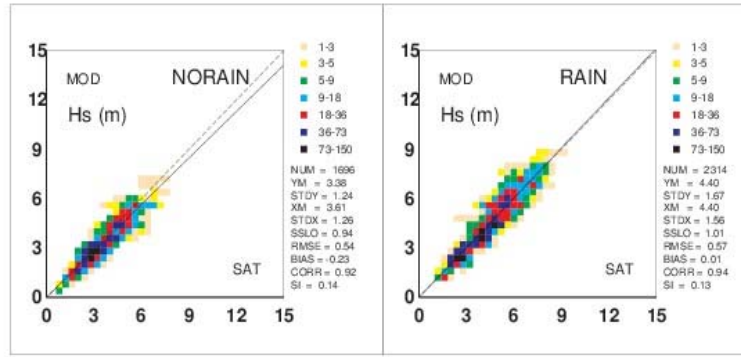

Figure 3 – As in Figure 2, but comparison with altimeter measured significant wave heights in non-rainy and rainy areas.

Figure 2 shows two scatter diagrams comparing the results from the EXP(eriment, i.e. with the zero-ed spectral tail depending on the rain rate) versus MOD(el, i.e. the standard) approach. The left panel shows the results for the non-rainy areas, the right RAIN one those for the rainy areas (in space and time). In each diagram the dash line shows the 45° perfect fit, the continuous line the through the origin best-fit to the data (symmetric slope, evaluated as $(\sum y^2/\sum x^2)^{1/2}$). The color of the different pixel indicates the respective number of data (see the side scales). The slopes are respectively 1.02 and 1.04 in the non-rainy and rainy areas. These figures are significant at more than 99% level. In Figure 3, we show the similar comparison for $H_s$ against altimeter data. While in the non-rainy areas the model underestimates 6%, in the rainy areas there is a 1% excess (again this difference is significant at more than 99% level). Given the no-rain - rain differences in the two figures, we conclude that the wave height results support an enhanced wave generation in rainy conditions. We stress that the significance of these results is not in the more or less good fit with the measured data, but in the difference between dry and rainy areas. Rain is not associated to a particular zone, but it is widely distributed throughout the overall area bordered in Figure 5, implicitly affecting also the non-rainy zones. To further check the significance of the results in Figure 3 we have repeated the analysis for each one of the seven smaller areas, marked 1 to 7, in Figure 5. Stressing again the relevance of the differences between non-rainy and rainy areas, all the corresponding best-fit slope differences are positive, varying between 1 and 9%, these ones too significant at more than 99%. Higher wave heights in rainy zones are coherent with our original physical hypothesis.

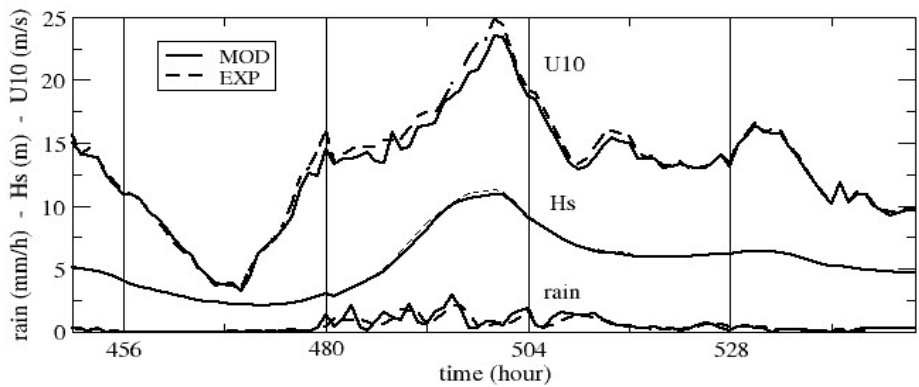

Figure 4 – Four day time series of model wind speeds, significant wave heights and rain rates at a position in the North Atlantic Ocean during December 2009. Continuous lines show the regular model results, the dash ones taking the reduced surface roughness under rain into account.

Having seen the general statistics, it is instructive to look at a time series of rain, wind and waves at a location during a rainy stormy event. Figure 4 shows the time evolution of the three quantities. The continuous and dash lines refer to the MOD and EXP runs respectively. We see the increased (with respect to MOD) wind speeds and significant wave heights during the period rain is present. Note that, while the wind increase mostly represents a local effect, waves derive also from the wind action during the previous hours. Although apparently limited, the half a meter peak $H_s$ difference is significant.

Although we did not investigate the matter further, we found that a different surface wind speed implies effects also on the upper layers of the atmosphere and in particular on the speed with which a meteorological system, typically a cold front, moves. This was verified plotting (not shown) the EXP-MOD rain rate differences across a front. Two parallel systems of large values were found, respectively positive and negative, consequent to the different position of the front in the EXP and MOD simulations. The suggestion is that taking rain-on-wave effect into account may change the propagation speed of the meteorological fronts.

To have a more comprehensive idea of the distributed effect on wave height, we have plotted in Figure 5 the differences EXP-MOD (respectively Exp74 and Exp73 in the title of the figure) at 09 UT of 10 December 2009. The isolines are at 10 cm interval. There is a wide distribution of positive differences, with values larger than 30 cm.

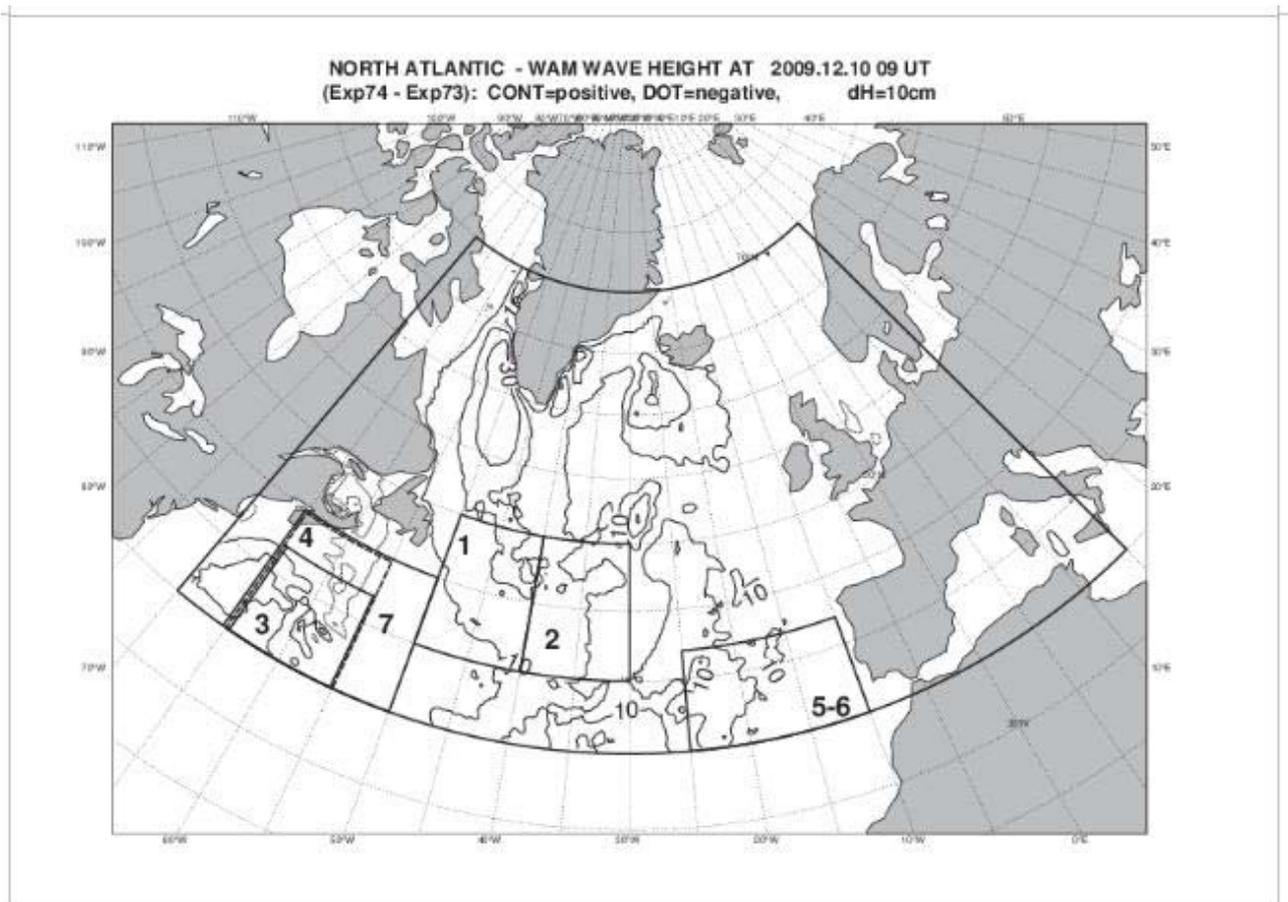

Figure 5 – North Atlantic area considered for the present analysis. The borders are 35° and 70° North, 70° West and 15° East. Areas 1 to 7 used for local statistics. Isolines (positive continuous, negative dashed) show

the significant wave height differences between the experimental and the regular runs at 09 UT 10 December 2009. Isolines at 10 cm interval. Areas with more than 30 cm difference are present.

We conclude this comparison by plotting in Figure 6, respectively for wind speed and significant wave height, the EXP-MOD percent differences as a function of the MOD values. The colors (scales on the right of each panel) code the number of events in each pixel. Within a relatively large scatter of the data, reflecting the distribution around the best-fit lines as in Figure 2 and Figure 3, there is a marked tendency towards positive, i.e. larger EXP, values. In our interpretation, much, if not most, of the scatter is, as just discussed, associated to slightly different positions of the fronts when the reduced surface roughness leads to different conditions in the atmospheric boundary layer.

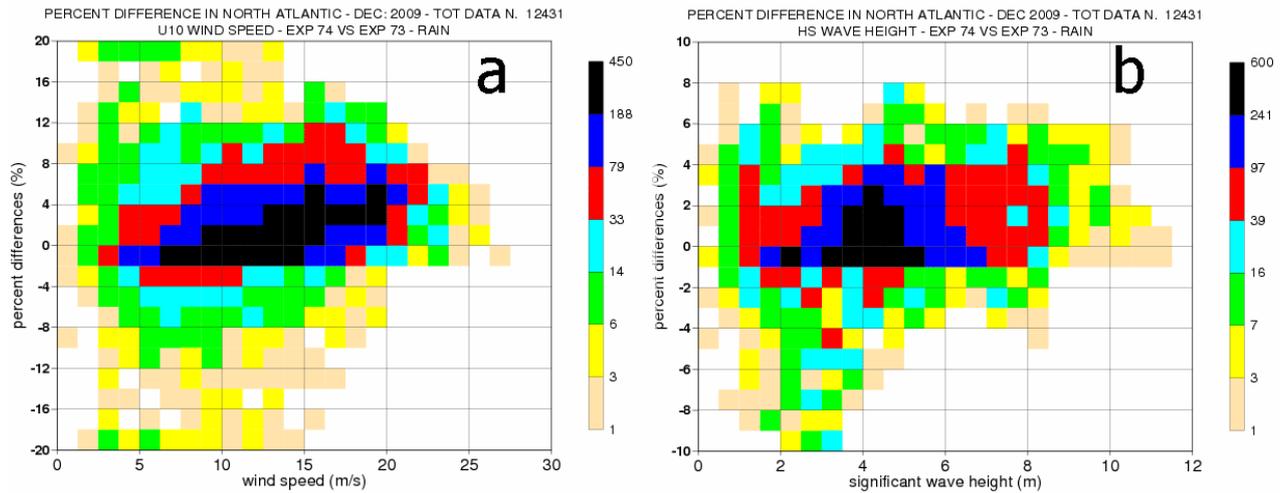

Figure 6 – Percent differences between the experimental run with reduced surface roughness under rain (EXP74) versus the regular one (EXP73). a) for wind speed, b) for significant wave height. The colors (scale on the right of each panel) code the number of occurrences in each pixel.

Having shown the assumed effect of rain in "calming the sea", hence on wind speed and on the overall wave height, it is now time to look for a thorough verification of the work hypothesis. This is the subject of the next section.

**5 – How a work hypothesis is not supported by the results**

In the previous sections, we have followed a logical course of action framing a numerical experiment to verify if our hypothesis of stronger wind speeds and higher wave heights in rainy conditions is true. From this perspective, the results have been positive in that indeed $U_{10}$ and $H_s$ appear to be, marginally but significantly, larger when rain is present. However, in a way this could have been taken for granted in that, once the hypothesis had been formulated and coded in the numerical model, the consequences followed in a logical sequence. Rather, it was a matter of quantifying the implications. Granted the approximation with which we have zero-ed the tail of the wave spectrum, the implications have turned out not large, but appreciable, also relevant given the present asymptotic (to measured data) performance of the operational results.

The next step is to verify if what we have found in our experiments is indeed present also in the daily data. This can be done analyzing the operational results versus measured data. If our hypothesis is true, wind speed and wave heights should have a tendency to be larger in rainy areas. Because this aspect is not present in the operational model, we should find different statistics for its performance in rainy and non-rainy areas. More specifically, when compared to measured data, the operational results should have lower best-fit slopes in rainy areas.

For the verification we have considered both altimeter ($H_s$) and buoy ($U_{10}$, $H_s$) data. Scatterometer and altimeter wind speeds have been excluded for the non-reliability of their signal in rainy areas, at least within the accuracy required for our verification. Because altimeter wave heights are assimilated into the ECMWF analysis, we have instead used the up to 12 hour forecasts. This has the further advantage that the model data are available at three hour interval (instead of the six for the analysis). To distinguish between rain and no-rain conditions, we have considered, for each time and location, the local hourly average amount of rain during the last three hours. Thinking of the integrated effect of wave generation, we have also used the last six hours tracking back where the local wind sea was during the previous hours. The two results, for three and six hour rain, are very similar. Only the three hour results are shown here.

The results of the comparison with North Atlantic buoy data, extended to the full 2009-2013 period, are summarized in Figure 7, upper panel for the (symmetric) slope, lower panel for the scatter index SI (rms error/mean measured value). Buoy wind data have been transformed to ten meter height assuming neutral stability conditions. The discrete values on the horizontal scales separate the various ranges of rain rate considered. The corresponding slope and SI results are plotted at the center of each range. The 0. point corresponds to the "no rain" cases. The far right RAIN point is for all the cases the rain rate is greater than 2.5 mmh$^{-1}$. The horizontal lines summarize all the results for rain rate > 0.1 mmh$^{-1}$.

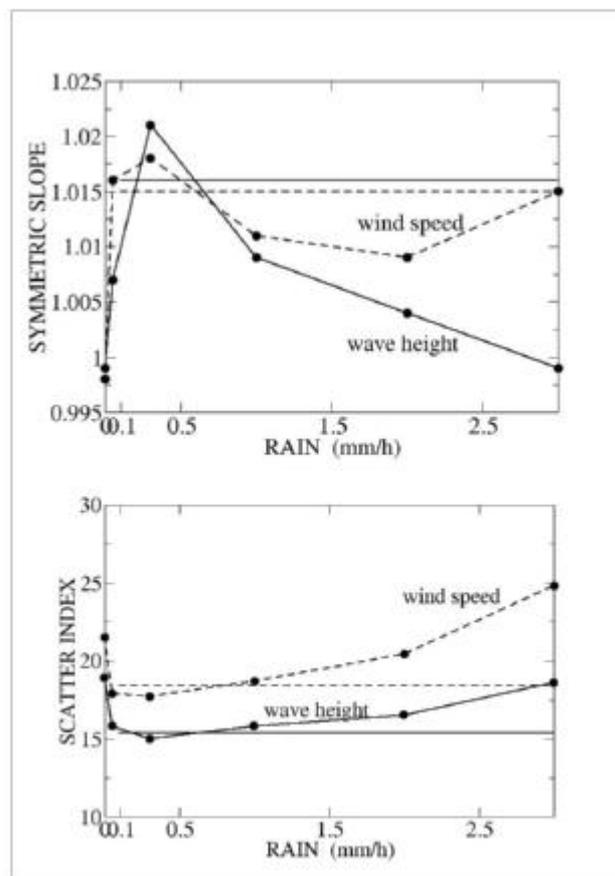

Figure 7 – Intercomparison between ECMWF operational model results and buoy recorded data. Both wind speeds and significant waves heights are considered. The overall area is shown in Figure 5. The period is 2009-2013. Upper panel for the symmetric slope, lower panel for the scatter index SI .The discrete values on the horizontal scales separate the various ranges of rain rate considered. The corresponding slope and SI results are plotted at the center of each range. The 0. point corresponds to the "no rain" cases. The upper

RAIN point is for all the cases the rain rate is greater than 2.5 mmh$^{-1}$. The horizontal lines summarize all the results for rain rate > 0.1 mmh$^{-1}$.

With the exception of the slope for the heaviest rains (soon to be commented about), it is clear that, contrary to our expectations, the model shows larger than measured $U_{10}$ and $H_s$ values when in rainy areas. The reliability of the results decreases for high rain rates because of the limited number of cases. This is clearly shown by the vertical position of the horizontal lines summarizing all the rain cases and close to the 0.1-0.5 mmh$^{-1}$ value. A well definite result comes from the scatter index where for both wind and waves the values increase with increasing rain. This suggests something is going on and that conditions are different under the rain.

The results are more definite for the altimeter wave heights (see Figure 8) where, consistently with Figures 2 and 3, we focus on the rainy 2009. Independently of the general slight underestimate by the model, the relevant point is that in rainy conditions the model provides higher wave heights with respect to the non-rainy cases. The difference is 4%, significant at more than 99%. Because the operational model does not consider the rain effect on wave generation by wind, the only conclusion is that in the ocean, again contrary to our expectations, the actual wave heights are lower when it is raining. This result holds for both December only (more and more intense rains) and for the full year. Indeed a corresponding analysis for different rain rates confirms (but with much less data in the high rate range) that the difference increases with more intense rain.

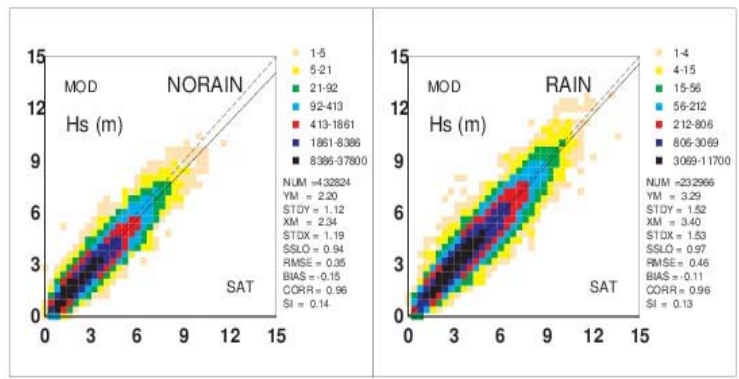

Figure 8 - Intercomparison between ECMWF operational model results and altimeter measured significant wave heights. The overall area is shown in Figure 5. The period is full year 2009. Left panel non-rainy, right panel rainy conditions. The colors identify, in a geometric scale, the number of data per pixel (see side scale). Also the overall statistics is reported.

This was a rather unexpected result that prompted a thorough verification of our procedure and of the analysis model versus measurements. The confirmation of the results shown in Figure 8 led to the conclusion that, for given forcing atmospheric conditions, wind waves tend to be lower in rainy areas. This requires a keen discussion that is the subject of the next, final section.

**6 – The search for an explanation**

Given that our results contradict the experimental evidence, clearly we need to look for the missing physics in our approach. We stress again that all our reported statistics, exemplified in the intercomparisons in Figures 2, 3 and 8, are very robust. We double-checked their significance using both the boot-strap and the jack-knife approaches (see, e.g., *Edgington*, 1995). All the results turned out significant at better than 99%.

We start with the simple, well acknowledged evidence that "rain calms the sea". While, as already specified, with this expression sea-men really mean the disappearance of the breaking crests, it is correct to try to quantify how much the rain can effectively dampen the sea. Practically, to obtain measurable results within a laboratory distance, all the related experiments had to work with very high rain rates, well above what experienced in the sea (with the exception of extreme cases). Also very short wave lengths had to be considered. Given the strong sensitivity of the attenuation to the wavelength, these results, although interesting from the physical point of view, are not very useful to evaluate the attenuation in the field main wave regime. For this we have reported the results of the theoretical approach of Le Mehaute' and Khangaonkar (1990, LMK) who again worked with substantial rain rates. With 50 mmh$^{-1}$ they found a $H_s$ attenuation of 38, 5 and 0.5 %h$^{-1}$ for 1, 10, 100 m wavelength respectively. Assuming these figures hold also when composed in a spectrum, considering, e.g., a 10 s peak period JONSWAP one, a quick estimate suggests an overall Hs decrease of 1.2%h$^{-1}$. Extrapolation (that we expect non-linear) of these figures to the more common 10-30 mm h$^{-1}$ of a cold front and to wave period of 10 s or more typical of the North Atlantic winter suggests a figure less than 0.5%h$^{-1}$, i.e. that rain is not directly, or the main, responsible (factor) for the lower wave heights in rainy areas we have considered.

Nevertheless we cannot neglect the perception that after a sustained rain we perceive (measurements are scarce in these conditions) that the wave heights are somehow lower. However, the key point we need to stress is that the disappearance, or decreasing number, of whitecaps under a strong rain is practically an instantaneous process (matter of seconds). The steep waves in Figure 1 are suggestive in this respect. The picture was taken in the Tahiti area during a squall. Being a localized event, model data, referred to a wider area, do not represent the specific conditions. The people onboard reported wind speed larger than 15 ms$^{-1}$ and significant wave height higher than 1 m. As also evident from the picture because of wave steepness, dominant waves were locally generated. Rain was reported as "very heavy". One of us (L.C.) has recently witnessed such an event during a wave measuring campaign on the ISMAR oceanographic tower (Northern Adriatic Sea, 16 meter depth, 15 km offshore the Venice coast). Figure 9 pictures the situation, with $H_s$=1.4 m, at three minute difference, before and soon after the onset of the downpour (32 mmh$^{-1}$ recorded on board). No marked variation of wind (about 14 ms$^{-1}$) was evident from the local record. Videos for the two situations are available as well. Clearly some different basic physics is at work.

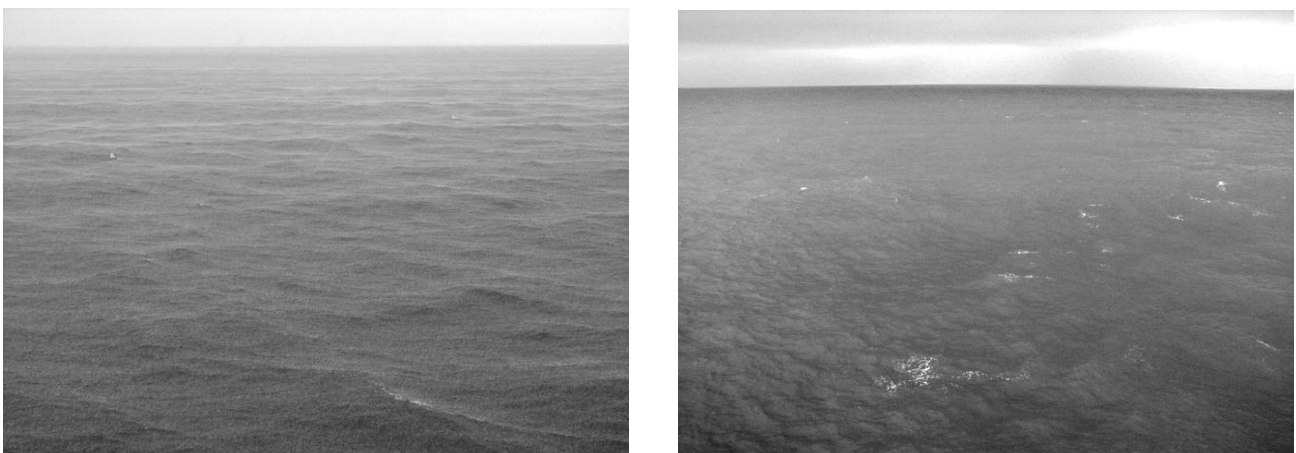

Figure 9 – Sea surface and white-capping distribution during (left) and just before (right) a violent and sudden downpour (32 mmh$^{-1}$). The two pictures have been taken at less than three minute distance. Significant wave height 1.4 m, wind speed 14 ms$^{-1}$. Oceanographic tower of ISMAR, 16 m depth, 15 km offshore Venice, Italy.

Following LMK, some physical intuition and some keen experiments as by *Peirson et al.* (2013), it is clear that rain acts mainly and more effectively on the tail of the spectrum. It is then natural to assume that what we see in a downpour, i.e. the disappearance of the whitecaps, is related to the different situation in the tail. This is supported by the results of Hwang and Wang (2004) suggesting that a strong signature of wave breaking is found in the 0.16 to 2.1 m wavelength scale. The related distribution in the field is not random, but, as direct evidence strongly suggests, mostly associated to the crests of the dominant waves.

A stronger effect turns out to concern the key element of any sea storm, i.e. the wind input to waves. The basic theory was formulated by Miles (1957, plus a later sequence of papers), then refined by Janssen (1989, 1991) to consider the feed-back of waves on the driving wind field. The theory stands on three basic concepts, those of roughness length $z_0$ (i.e. the height close to the surface where the wind speed is null), of a logarithmic vertical profile of the wind speed, and of critical height $z_c$ as the one where wind speed equals the phase speed of the wave component we consider acting upon. The major drag of the wind profile, hence a large momentum flux from wind to waves, is given by the tail of the spectrum with its low but steep waves, hence the low values of $z_0$. If these waves are flattened by rain $z_0$ increases (now related to longer and higher waves), and the logarithmic profile, so to say, relaxes, with a less steep increase of the wind speed close to the surface. In turn this implies a substantial increase of the critical height $z_c$. This has a dramatic effect on the input by wind to the considered component, because the momentum transfer to waves is roughly proportional to $\exp(-2kz_c)$ with k the wave number (see the discussion by Lighthill, 1962). Indeed the Miles-Janssen mechanism is effective only for low values of the critical height with respect to the wavelength. This makes the input to the bulk of the spectrum extremely sensitive to the condition of the tail. Following the above argument a heavy rain, via its effect on the tail, indirectly affects the whole generation process. Because during generation most of energy input by wind (>90%) is immediately lost as white-capping (see, e.g., Holthuijsen, 2007), cancelling the tail leads also to an immediate decrease, in the extreme the disappearance, of white-capping.

So the tail of the spectrum acts on surface breaking in two ways: via the direct influence of the tail on the breaking process, and via the possible drastic decrease of input by wind. In this second aspect it is worth mentioning that there is a feed-back process at work. In 1976 Banner and Melville pointed out that much, if not most, of the momentum input by wind to waves does not take place as a smooth continuous process. Rather, it happens in bursts, mostly in connection with the breakers, or white-caps, that characterize the crests of a sea under the vigorous action of wind. *Kudryavtsev and Makin* [2001] further explored this possibility evaluating the form drag associated to the presence of breakers. *Babanin et al.* [2007], following their AUSWEX experiment, concluded that the presence of breakers implies a significant phase shift in the local wave-coherent surface pressure. This produces a wave-coherent energy flux from wind to waves with a mean value twice the corresponding energy flux to the non-breaking waves. So the decreased input by wind because of rain is further decreased because of the reduced white-capping.

Another point to be discussed is the modification of the non-linear interactions balance once the tail of the spectrum is flattened. Still sticking for simplicity to the classical JONSWAP spectrum of a generating sea, we know (see for instance the detailed and keen analysis by Young and Van Vledder, 1993) that there is a flow of energy from the central part of the spectrum towards higher frequencies where it is dissipated by white-capping. If we zero the tail, the non-linear interactions will tend to reestablish the "correct" spectral shape. In practice there will be an increased flow of energy from the central part of the spectrum towards the tail. A first-hand estimate is obtained estimating the energy cancelled by rain and the consequences on the overall energy and the associated significant wave height. Using again the figures provided by LMK, we obtain a figure close to 1%, far from the values required to justify the overall situation. A more precise estimate has been obtained with the full evaluation of the related Boltzmann integral using ExactNL. Also this result indicates that the flow of energy towards the tail cannot justify a rapid appreciable decrease of the significant wave height. In any case the test has also highlighted how approximate is in this case the Discrete Interaction Approximation (DIA, Hasselmann et al., 1985) used every day in operational modeling.

With respect to the direct action of rain on wind waves, we need to consider (see LMK) that, if sufficiently intense and falling at an angle, rain can add energy and momentum to waves. A similar argument concerns the foaming crests under a strong wind when their tearing leads to large, rapidly accelerated water drops impinging on the back side of the previous wave. The argument is subtle because the energy and momentum of the flying rain or drops are extracted from the wind. In the case of rain Manton (1972) has roughly quantified the loss of wind speed at the $10^{-3}$ level. While the overall energy of the system (rain, drops and wind) remains the same, we should possibly consider that their direct impinging on the previous wave can be a more efficient way of transferring energy from air to waves.

Two things need to be pointed out about the effect of rain or torn drops on waves. The nice machine we have assembled in a numerical wave model assumes a clean well defined separation surface between air and water. However, when we go to very high wind speeds, the obvious examples being hurricanes, the separation surface begins to lose its meaning, substituted by a layer of "foaming material", troughs full of foam, and so on. The physics of this interface, still a subject of valuable studies (see, e.g., Soloviev and Lukas, 2010), is different, and it is amazing that our wave models still manage to produce valuable results also in these conditions. Remaining on more solid (or liquid) ground, in this paper we deal with less extreme conditions exploring the direct effect of rain on wave generation.

Focusing on the meteorological aspect of rain, it is relevant that the horizontal momentum of rain is presently not considered in meteorological modeling (Anton Beljars, personal communication). More in general, the water cycle, and more specifically the rain rate, is still open to improvement in meteorological modeling (see, e.g., http://old.ecmwf.int/publications/library/do/references/show?id=90901). In particular the ECMWF model has a tendency to drizzle, i.e. to overpredict very low rain rates. For this reason we have excluded this lower range from our analysis. Another important aspect is that, independently of the model accuracy, a once-an-hour or once-every-three-hour integrated rain datum is likely to smooth or average the truth. Rain comes often in bands, a very patchy process, both in space and time, with rapid and alternating variations. If the process considered is non-linear, to apply the overall average to the process under consideration may lead to a different result. On the other hand this is the information we have at disposal and have to work with. In a way the situation is similar to wind gustiness in wave modeling. *Abdalla and Cavaleri* [2002] showed that, following the standard approach to the generation by wind, a gusty wind leads to higher wave heights than using its average value. At least on the average, the problem was partially solved having an estimate of the level of gustiness and, knowing the related physics, deriving a parameterized estimate of the implication [*Janssen*, 2008]. Somehow, having an estimate of the rain patchiness, the same approach could be followed for the implications for wind and waves.

On a parallel criticism on wave modeling, we need to realize that white-capping is still the least understood process in wave modeling, for this reason often used as the tuning knob of the system. A somehow opposite, but solid physical approach to the process of white-capping was given by *Banner et al.* [2002] who pointed out how the single crests may reach breaking conditions as a consequence of the energy convergence while reaching the top of the enclosing envelope. However, at the model scale much simpler approaches are used (see, e.g., *Bidlot et al*, 2012 and *Ardhuin et al.*, 2010) based on the shape of the spectrum, but not very sensitive to the presence of the tail. Unluckily no detailed wave spectra exist taken under rainy conditions, certainly not in the frequency range high enough to observe the behavior of the tail in these conditions. Physically, based on theory, experiments and intuition, it is reasonable to expect, while the rain rate increases, a progressively larger section of the tail to be affected, starting from the high frequency end and gradually extending towards the lower frequencies. In turn this will affect the white-capping more and more, again starting from the crests of the shorter waves, and progressively extending to the ones of the longer and dominant waves (as it is the case in Figure 1).

In summary, the picture we have reached of the situation is the following. Rain, if sufficiently intense, and as done with oil especially in the past, dampens the high frequency part of the wave spectrum. With increasing rain rate the dampening begins in the capillary range extending progressively, but with progressively longer time scale, to lower frequencies. The smoother surface leads to a lower friction of wind on the surface, in so doing reducing the slow down due to wave coupling and their growth rate. However, we find also an increase of the roughness length and of the critical height in the Miles-Janssen generation process. This strongly reduces the input by wind on the whole spectral range. Because most of the wind input to a growing sea is immediately lost as white-capping, this implies also a much reduced breaking rate. With time passing, and rain continuing, the height and steepness of the waves tend to decrease, but it is important to realize that the decreased generation and number of breakers are immediate effects at the onset of rain (if sufficiently intense). The two facts, reduced generation and breaking, loop on each other because breaking enhances wind input via the possible detachment of the surface layer at the crest.

A practical problem for the proper consideration of the above process in wave modeling is the approximation still present in rain forecast, both because of its patchiness and the uncertainties that still characterize the water cycle in meteorological modeling. For the former a statistical approach similar to what done for wind gustiness can be a possible solution.

Because at the onset of a heavy rain the white-capping disappears vary rapidly (matter of seconds or little more), the wave spectrum does not change drastically across this short transient. This suggests that the present white-capping quantification in wave modeling, related to the distribution of energy in the spectrum, is not fully correct. Historically, the evolution of wave field (in deep water) has been conceived as the result of three different processes: wind input, non-linear interactions, and white-capping. We suggest that wind input and white-capping should be considered as a single process. Indeed, in engineering terms, the present approach may appear difficult and illogical. Wave growth is evaluated as a minor difference, almost two orders of magnitude smaller, between two large quantities, wind-input and white-capping, independently evaluated, the latter one still considered not properly known (hence its use as a tuning knob). This is prone to errors. Given the physical connection between the two processes, we suggest that a new approach, based on a single view of these two strongly connected, but presently separate, processes, is the way to follow in the future. Remarkable indications, as the ones by Banner and Melville (1976) and Kudryavtsev and Makin (2001), had already shown the reciprocal influence of white-capping and generation. However, although in an undefined way for which we beg the reader pardon, we feel that a more unified concept and definition, physical and numerical, is required. White-capping and generation, both not existing without the other one. Both crucial for all the exchanges (mass, energy, heat, momentum, etc.) between atmosphere and ocean. So crucial not only for the single storm, but also for the Earth climate. An updated, more solid and complete approach is badly required.

**Acknowledgements**

Model data used in this work have been retrieved from the ECMWF archive system. Following ECMWF data policy, the data are not directly available, but they can be requested to one of the authors: luigi.cavaleri@ismar.cnr.it, luciana.bertotti@ismar.cnr.it, jean.bidlot@ecmwf.int.
Luciana Bertotti and Luigi Cavaleri have carried out this work thanks to the support of the EU funded projects Field-AC (FP7-SPACE-2009-1) and MyWave (SPA.2011.5-03). They are also thankful for the possibility to use the ECMWF computer system and facilities under the local

Special Projects plan. Discussions with Philippe Lopez, Saleh Abdalla, Peter Janssen and Piero Lionello are much appreciated. The tests with the ExactNL have been carried out by Gerbrant van Vledder whom we warmly thank for his kind availability and competence. Figure 1 has been kindly made available by Ginni Callahan. We acknowledge the useful comments by three anonymous reviewers on the previous version of this paper, comments that led to a much more complete view of the problem.

## References

Abdalla, S. and L.Cavaleri, (2002), Effect of wind variability and variable air density in wave modeling, *J.Geoph.Res.*, Vol.107, No.C7, 17-1/17-17.

Ardhuin, F., W.E.Rogers, A.V.Babanin, J.-F.Filipot, R.Magne, A.Roland, A.J.van der Westhuysen, P.Queffeulou, J.-M.Lefevre, L.Aouf, and F.Collard, (2010), Semi-empirical dissipation source functions for ocean waves: Part I, definitions, calibration. *J. Phys. Oceanogr.*, 40(9), 1917–1941, doi:10.1175/2010JPO4324.1

Babanin, A.V., M.L.Banner, I.R.Young, and M.A.Donelan, (2007), Wave-follower field measurements of the wind-input spectral function. Part III: parameterization of the wind-input enhancement due to wave breaking, *J. Phys. Oceanogr.*, **37**, 2764-2775.

Banner, M.L.. and W.K.Melville, (1976), On the separation of air flow over water waves, *J. Fluid Mech.*, **77**, part 4, 825-842.

Banner, M.L., J.R.Gemmrich, and D.M.Farmer, (2002), Multiscale measurements of ocean wave breaking probability. *J. Phys. Oceanogr.*, **32**, 3364–3375.

Beya, J., W.Peirson and M.Banner, (2010), Attenuation of gravity waves by turbulence, Proc. 32$^{nd}$ Int. Conf. Coastal Eng., Shanghai, China.

Bidlot J.-R., (2012), Present status of wave forecasting at ECMWF. Proceeding from the *ECMWF Workshop on Ocean Waves*, 25-27 June 2012. ECMWF, Reading, U.K., http://old.ecmwf.int/publications/library/do/references/list/201210251

Bidlot J.-R., J.-G. Li, P. Wittmann, M. Faucher, H. Chen, J.-M, Lefevre, T. Bruns, D. Greenslade, F. Ardhuin, N. Kohno, S. Park and M. Gomez, (2007), Inter-Comparison of Operational Wave Forecasting Systems. Proc. 10th International Workshop on Wave Hindcasting and Forecasting and Coastal Hazard Symposium, North Shore, Oahu, Hawaii, November 11-16,2007.
http://www.waveworkshop.org/10thWaves/ProgramFrameset.htm

Braun, N., M.Gade and P.A.Lange, (2002), The effect of artificial rain on wave spectra and multi-polarisation X band radar backscatter, *J. Remote Sens.*, **23**, 4305-4322, **DOI:** 10.1080/01431160110106032

Chen, G., B.Chapron, J.Tournadre, K.Katsaros and D.Vandemark, (1998), Identification of possible wave damping by rain using TOPEX and TMR data, *Remote Sens. Environ.*, **68**, 40-48.

Edgington, E.S., (1995), *Randomization tests*, M.Dekker, N.Y., 409p.

ESA, 2013, Rain cells over the ocean, Earthnet Online,
earth.esa.it/applications/ERS-SARtropical/atmospheric/rains/intro/

Hasselmann, K., (1962), On the non-linear energy transfer in a gravity wave spectrum, part 1: general theory, *J. Fluid Mech*, **12**, 481

Hasselmann, S., K.Hasselmann, J.H.Allender and T.P.Barnett, (1985), Computations and parameterizations of the nonlinear energy transfer in a gravity wave spectrum, part 2: Parameterization of the nonlinear energy transfer for application in wave models, *J. Phys. Oceanogr.*, **15**, 1378-1391.

Hwang, P.A., and D.W.Wang, 2004, An empirical investigation of source term balance of small scale surface waves, *Geoph.Res.Lett.*, **31**, L16301, 5pp, doi;10.1029/2004GL020080, 2004.

Janssen, P.A.E.M., (1989), Wave-induced stress and the drag of air flow over sea waves, *J. Phys. Oceanogr.*, **19**, 745-754, doi: http://dx.doi.org/10.1175/1520-0485(1989)019<0745:WISATD>2.0.CO;2

Janssen, P.A.E.M., (1991), Quasi-linear theory of wind wave generation applied to wave forecasting, *J. Phys. Oceanogr.*, **21**, 1631-1642, doi: http://dx.doi.org/10.1175/1520-0485(1991)021<1631:QLTOWW>2.0.CO;2

Janssen, P.A.E.M., (2004). *The interaction of ocean waves and wind.* Cambridge University Press, 300p.

Janssen, P.A.E.M., (2008), Progresses in ocean wave forecasting, *J. Comp. Physics,* **227**, 3572-3594, http://dx.doi.org/10.1016/j.jcp.2007.04.029

Komen, G.J., L.Cavaleri, M.Donelan, K.Hasselmann, S.Hasselmann and P.A.E.M.Janssen, (1994), *Dynamics and Modeling of Ocean Waves*, Cambridge University Press, 532p.

Kudryavtsev, V.N. and V.K.Makin, (2001), The impact of air flow separation on the drag of the sea surface, *Bound.-Layer Meteorol.*, 98, 155-171.

Le Méhauté, B. and T.Khangaonkar, (1990), Dynamic interaction of intense rain with water waves, *J. Phys. Oceanogr.*, **20**, 1805-1812, doi: http://dx.doi.org/10.1175/1520-0485(1990)020<1805:DIOIRW>2.0.CO;2

Manton, M.J., (1973), On the attenuation of sea waves by rain, *Geoph. Fluid Dynam.*, **5:1**, 249-260,**DOI:**10.1080/03091927308236119

Miles, J.W., (1957), On the generation of surface waves by shear flows, *J. Fluid Mech*, **3**, 185-204.

Naeije,M., R.Scharroo, E.Doornbos, and E.Schrama, (2008). Global Altimetry Sea-level Service: GLASS, Final Report. NIVR/DEOS publ., NUSP-2 report GO 52320 DEO, 107p

Peirson, W.J., J.Beya, M.Banner, J.Peral and S.Azarmsa, (2013), Rain-induced attenuation of deep-water waves, *J. Fluid Mech.*, **724**, 5-35.

Pierson, W.J. and W.Marks, (1952), The power spectrum analysis of ocean-wave records, *Trans. Amer. Geoph. Union*, **33**, doi: 10.1029/TR033i006p00834 issn: 0002-8606

Poon, Y.-K., S.Tang and J.Wu, (1992), Interactions between rain and wind waves, *J. Phys. Oceanogr.*, **22**, 976-987, doi: http://dx.doi.org/10.1175/1520-0485(1992)022<0976:IBRAWW>2.0.CO;2

Reynolds, O., (1875), Rain-drops on the sea, *Popular Science Monthly*, Vol.6.

Reynolds, O., (1900), On the action of rain to calm the sea, *Papers on Mechanical and Physical Subjects*, Vol.1, Cambridge University Press, 86-88.

Soloviev, A. and R.Lukas (2010), Effects of bubbles and sea spray on air-sea exchange in hurricane conditions, *Boundary-Layer Meteorol.*, **135**, 365-376, DOI:10.1007/s10546-010-9505-0

Simmons, A.J. and J.K.Gibson, (2000), The ERA-40 Project Plan, ERA-40 Project Report Series n.1, ECMWF, Reading, 62p.

Tsimplis, M.N., (1992), The effect of rain in calming the sea, *J. Phys. Oceanogr.*, **22**, 404-412, doi: http://dx.doi.org/10.1175/1520-0485(1992)022<0404:TEORIC>2.0.CO;2

Tsimplis, M.N. and S.A.Thorpe, (1989), Wave damping by rain, *Nature*, **342**, 893895.

Weissman, D.E., B.W.Stiles, S.M.Hristova-Veleva, D.G.Long, D.K.Smith, K.A.Hilburn and W.L.Jones, (2012), Challenges to satellite sensors of ocean winds: addressing precipitation effects, *J. of Atm. and Ocean. Tech.*, **29**, 356-374.

Young.I.R., and G.Ph.Van Vledder, 1993, A review of the central role of nonlinear interactions in wind-wave evolution, *Phil.Trans.R.Soc.Lond*. A, **342**, 505-524.

**Figure captions**

Figure 1 – Wind sea in rainy conditions (courtesy of Ginni Callahan). Environmental conditions (on-board estimate): wind speed 15 ms$^{-1}$, significant wave height > 1 m, mainly wind sea, very heavy rain.

Figure 2 – EXPerimental run with reduced surface roughness under rain versus the regular MODel results. Surface wind speeds over the North Atlantic Ocean (see Figure 5 for the considered area). Left panel results in the non-rainy areas, right panel in rainy areas. Colors represent, in geometric progression, the number of cases in each pixel (see the side scale). Also the overall statistics is reported.

Figure 3 – As in Figure 2, but comparison with altimeter measured significant wave heights in non-rainy and rainy areas.

Figure 4 – Four day time series of model wind speeds, significant wave heights and rain rates at a position in the North Atlantic Ocean during December 2009. Continuous lines show the regular model results, the dash ones taking the reduced surface roughness under rain into account.

Figure 5 – North Atlantic area considered for the present analysis. The borders are 35 $^o$ and 70 $^o$ North, 70 $^o$ West and 15 $^o$ East. Areas 1 to 7 used for local statistics. Isolines (positive continuous, negative dashed) show the significant wave height differences between the experimental and the regular runs at 09 UT 10 December 2009. Isolines at 10 cm interval. Areas with more than 30 cm difference are present.

Figure 6 – Percent differences between the experimental run with reduced surface roughness under rain (EXP74) versus the regular one (EXP73). a) for wind speed, b) for significant wave height. The colors (scale on the right of each panel) code the number of occurrences in each pixel.

Figure 7 – Intercomparison between ECMWF operational model results and buoy recorded data. Both wind speeds and significant waves heights are considered. The overall area is shown in Figure 5. The period is 2009-2013. Upper panel for the symmetric slope, lower panel for the scatter index SI .The discrete values on the horizontal scales separate the various ranges of rain rate considered. The corresponding slope and SI results are plotted at the center of each range. The 0. point corresponds to the "no rain" cases. The upper RAIN point is for all the cases the rain rate is greater than 2.5 mmh$^{-1}$. The horizontal lines summarize all the results for rain rate > 0.1 mmh$^{-1}$.

Figure 8 - Intercomparison between ECMWF operational model results and altimeter measured significant wave heights. The overall area is shown in Figure 5. The period is full year 2009. Left panel non-rainy, right panel rainy conditions. The colors identify, in a geometric scale, the number of data per pixel (see side scale). Also the overall statistics is reported.

Figure 9 – Sea surface and white-capping distribution during (left) and just before (right) a violent and sudden downpour (32 mmh$^{-1}$). The two pictures have been taken at less than three minute distance. Significant wave height 1.4 m, wind speed 14 ms$^{-1}$. Oceanographic tower of ISMAR, 16 m depth, 15 km offshore Venice, Italy.